\documentclass[12pt,preprint]{aastex}

\shorttitle{Burst Spectra of EXO~0748$-$676}
\shortauthors{Cottam et al.}

\begin{document}


\title{The Burst Spectra of EXO~0748$-$676 during a Long 2003 XMM-Newton Observation}


\author{J. Cottam\altaffilmark{1}, F. Paerels\altaffilmark{2,3}, M. M\'{e}ndez\altaffilmark{3}, L. Boirin\altaffilmark{4}, W.H.G. Lewin\altaffilmark{5}, E. Kuulkers\altaffilmark{6}, J.M. Miller\altaffilmark{7}}

\altaffiltext{1}{NASA Goddard Space Flight Center, Astrophysics Science Division, Greenbelt, MD 20771} 
\email{jean.cottam@nasa.gov}

\altaffiltext{2}{Columbia Astrophysics Laboratory, Columbia University, 550 West 120th Street, New York, NY 10027} 

\altaffiltext{3}{SRON National Institute for Space Research, Sorbonnelaan 2, 3584 CA Utrecht, The Netherlands} 

\altaffiltext{4}{Observatoire Astronomique de Strasbourg, 11 rue de l'Universit\'{e}, 67000 Strasbourg, France}

\altaffiltext{5}{Kavli Institute for Astrophysics and Space Research, 70 Vassar Street, Cambridge, MA 02139}

\altaffiltext{6}{ISOC, ESA/ESAC, Urb. Villafranca del Casstillo, PO Box 50727, 28080 Madrid, Spain}

\altaffiltext{7}{Department of Astronomy, University of Michigan, Ann Arbot, MI 48109} 


\begin{abstract}
Gravitationally redshifted absorption lines from highly ionized iron have been previously identified in the burst spectra of the neutron star in EXO~0748$-$676.  To repeat this detection we obtained a long, nearly $600\, {\rm ks}$ observation of the source with XMM-Newton in 2003.  The spectral features seen in the burst spectra from the initial data are not reproduced in the burst spectra from this new data.  In this paper we present the spectra from the 2003 observations and discuss the sensitivity of the absorption structure to changes in the photospheric conditions. 
\end{abstract}



\keywords{binaries: general ---
stars: individual(\objectname{EXO~0748$-$676}) --- 
stars: neutron-stars --- 
X-rays: bursts --- 
X-rays: binaries} 


\section{Introduction}

In \citet{CPM} we identified discrete absorption features corresponding to electronic transitions in highly ionized iron in the burst spectra of the neutron star in EXO~0748$-$676.  
Using data acquired during the commissioning and calibration of the XMM-Newton observatory we compiled spectra from 28 type I X-ray bursts in EXO~0748$-$676.  
After accounting for the effect of circumstellar absorption we identified features originating in the thermal emission spectrum from the neutron star atmosphere.  
We identified an \ion{Fe}{26} $n=2-3$ absorption feature (from H-like Fe) in the average spectrum of the early phases of the bursts, and an analogous \ion{Fe}{25} $n=2-3$ absorption feature (from He-like Fe) in the spectrum of the late phases of the bursts.  
Both features exhibited an identical gravitational redshift of $z=0.35$.  Additional absorption structure observed between $25$ and $27\, {\rm \AA}$ during the late phases of the bursts was tentatively identified as due to \ion{O}{8} $n=1-2$ at the same redshift.  
A gravitational redshift of $z=0.35$ at the neutron star surface translates to a mass-to-radius ratio of $M/R = 0.152\, {\rm M_{\sun}/km}$ for the neutron star in EXO~0748$-$676.  This measurement provides an empirical constraint on the equation of state of dense, cold nuclear matter.  

In a subsequent analysis \citet{VS2004} succeeded in measuring the spin frequency for the neutron star in EXO~0748$-$676 from coherent modulations of the X-ray flux during the decaying portion of its X-ray bursts. The spin frequency, $45\, {\rm Hz}$, is much lower than measured in any other Low Mass X-ray Binary ({\it e.g.} \citet{Chak}).  
\citet{VS2004} show that the widths of the narrow line profiles that we observed in the XMM-Newton spectra are consistent with the spin frequency of the star for a neutron star radius between $9.5$ and $15\, {\rm km}$. The observed strength of the absorption lines implies that the line broadening is dominated by the Stark effect \citep{Paerels97,Bildsten}, which in principle allows for an independent measurement of the acceleration of gravity at the stellar surface.
\Citet{Chang06} performed a more detailed analysis of the observed line profiles including both the intrinsic line broadening and the rotational line broadening.  Since the two line broadening mechanisms are of very similar scale, they show that a direct measurement of the neutron star radius can not be made from the existing XMM-Newton data.  
\citet{Ozel} has used the measured value of the gravitational redshift, combined with an estimate of the stellar radius, to infer a mass for the neutron star of $M \geq 1.8 M_{\odot}$ for a stellar radius of $R \geq 12$ km in the coordinate frame of the neutron star.  

Because of the significance of these first results we requested additional XMM-Newton observations of EXO~0748$-$676 to repeat this measurement and extend the spectroscopic analyis.  
Almost $600\, {\rm ks}$ of XMM-Newton Director's Discretionary time was awarded.  In this paper, we will describe these new data and compare them to the 2000 data described in \citet{CPM} (hereafter CPM).  
We also requested Chandra observations of EXO~0748$-$676 to provide independent corroborating evidence for the presence of gravitationally redshifted absorption lines. A total of $300\, {\rm ks}$ of Chandra/HETGS data was acquired during the same time frame as the XMM-Newton data.  Those data will be presented in Paerels et al.\ (2007, in preparation).  

\section{Observations \& Data Reduction}

EXO~0748$-$676 was observed by the XMM-Newton Observatory \citep{XMM} on seven occasions during the fall of 2003 for a total exposure of $570\, {\rm ks}$.  The first four observations were conducted in sequential satellite orbits beginning on September 19, 2003.  
Three more observations were conducted beginning on October 21, October 25, and November 12, 2003.  The details of these observations are listed in Table \ref{tbl1}.  For each observation, data were acquired with all of the onboard instruments simultaneously.  
\clearpage
\begin{table} 
\begin{center}
\caption{XMM-Newton Observations of EXO~0748$-$676 in 2003 \label{tbl1}} 
\begin{tabular}{ccccccccc}
\tableline
\tableline 
Obs. ID & & Rev. & & Obs. Start & & Exp. & Bursts & Bursts \\ 
        & &      & &  (UT)      && (ks) &  EPIC/pn & RGS \\ 
\tableline 
0160760101 & & 0692 & & 09-19:13:12 & & 88.8  & 10  & 10 \\ 
0160760201 & & 0693 & & 09-21:13:13 & & 90.7  & 14  & 11 \\ 
0160760301 & & 0694 & & 09-23:10:17 & & 108.2 & 14  & 13 \\ 
0160760401 & & 0695 & & 09-25:17:05 & & 73.7  & 8   & 9 \\ 
0160760601 & & 0708 & & 10-21:09:38 & & 55.3  & 8   & 7 \\ 
0160760801 & & 0710 & & 10-25:18:55 & & 62.5  & 9   & 8 \\ 
0160761301 & & 0719 & & 11-12:07:59 & & 90.9  & 12  & 10 \\ 
\tableline 
\label{TabValues}
\end{tabular}
\end{center}
\end{table}
\clearpage
We are primarily interested in the high-resolution spectral data from the Reflection Grating Spectrometer (RGS), which cover the wavelength band from $\lambda \sim 5$ to $35\, {\rm \AA}$ ($E \sim 0.35$ to $2.5\, {\rm keV}$) with a resolving power of $\sim 300$ at $15\, {\rm \AA}$ \citep{RGS}.  
The RGS data were initially processed with the XMM-Newton Science Analysis Software (SAS) version 5.4.1, and post-processing was performed using SAS versions 6.0.0 and 6.5.0.  
For the final spectral analysis we reprocessed all data with SAS version 7.0.0.  In this analysis we used data from the European Photon Imaging Cameras (EPIC) only for the purposes of identifying the bursts.  
We used the EPIC/pn data \citep{EPICpn}, which were acquired in 'Small Window' mode to minimize CCD pile-up effects during the bright EXO 0748$-$676 bursts.  The EPIC/pn detectors cover the energy range of $\sim 0.15$ to $15\, {\rm keV}$.  A detailed analysis of the EPIC burst data is presented in \citet{Boirin}.   

\begin{figure}
\begin{center}
\includegraphics[angle=90,height=3.7in]{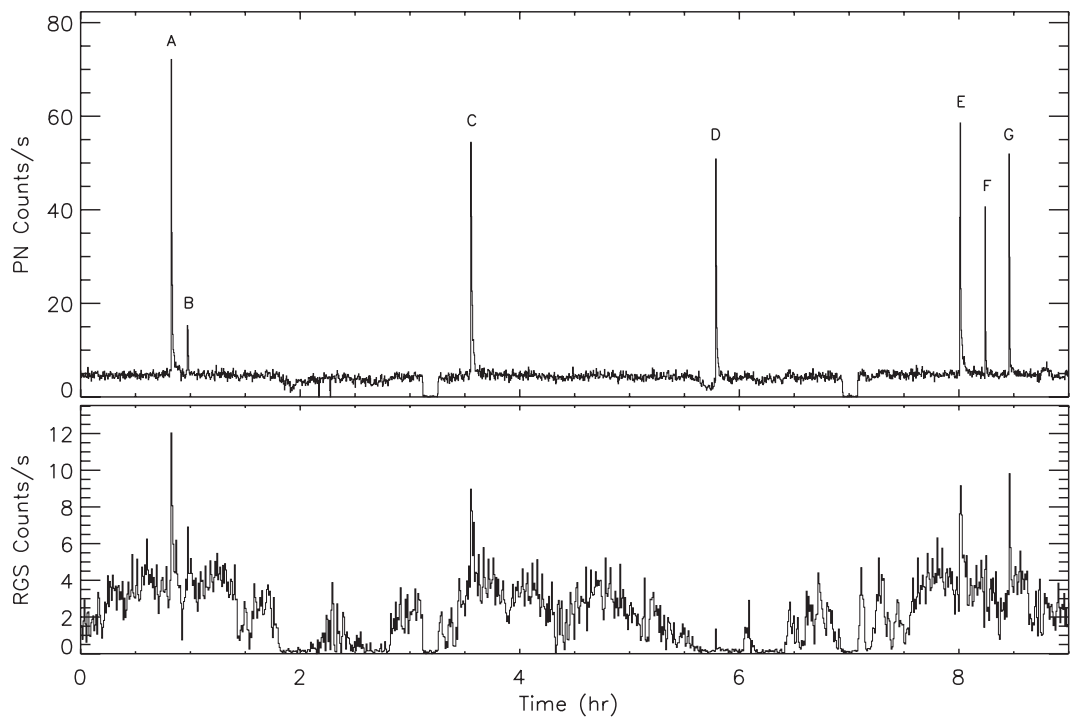} 
\end{center}
\vspace{-0.9cm}
\caption{Sample lightcurve (33 ksec, or 2.40 binary orbits of 3.82 hr) of EXO~0748$-$676 for EPIC/pn (top) and RGS (bottom), starting around 2003-09-22:03 (UT).  The EPIC/pn data cover the energy range of $5 - 10\, {\rm keV}$ and are binned at 16 seconds/bin.  The RGS data cover the energy range of $0.35 - 2.5\, {\rm keV}$ and are binned at 32 seconds/bin.  Two binary eclipses are visible
as is irregular dipping behavior.  Type I x-ray bursts are visible in both data sets, but are cleanest with the highest contrast in the EPIC/pn data.  The bursts labelled "B", "D" and "F" are not robustly identified in the RGS lightcurve.
}
\vspace{0.25cm}
\label{fig1} 
\end{figure}

We identified the type-I X-ray bursts using both the EPIC and RGS lightcurves. 
For the EPIC/pn we used only single and double events (patterns 0 to 4) and restricted the energy range to $5$ to $10\, {\rm keV}$ (see \citet{Boirin}).  For the RGS we used the standard $m=-1$ events over the full energy range of the instrument.  We defined the start of each burst as the time when the count rate increased above the local pre-burst persistent level by a factor of three, and the end as the time when the count rate decreased to the pre-burst persistent level.  
This is sometimes difficult to determine in the RGS data because of the low contrast of the bursts and the large variations in the persistent count rate caused by the larger effects of circumstellar absorption in the soft x-ray band (see Figure \ref{fig1}).  
We therefore used the EPIC/pn data to guide our search for bursts in the RGS lightcurve.   

The detailed burst characteristics from the EPIC/pn data are quoted in Table A.1 in \citet{Boirin}.  
In the EPIC/pn lightcurve, we identify 75 bursts.  We classified these as 34 single bursts, 13 double bursts and 5 triple bursts.  The burst duration ranges from 15 to 177 seconds for a cumulative burst exposure time of 6428 seconds.  
The peak intensity ranges from 12.2 to 376.4 counts s$^{-1}$ with an average of 195 counts s$^{-1}$ over the persistent count rate, which varied from 62 counts s$^{-1}$, down to 12 counts s$^{-1}$ during the deepest dips.  In the RGS lightcurve, we identify 68 bursts.  
One burst is only visible in the RGS lightcurve because of an offset in the times when the EPIC/pn and RGS instruments were turned on.  There are eight bursts identified in the EPIC/pn data that are not distinguishable in the RGS lightcurve \citep{Homan}.  The bursts cannot be easily classified as single, double, or triple bursts using the RGS lightcurve alone.  In some cases the second burst and in others the third burst in a series identified by the EPIC/pn as a triple burst can not be distinguished above the persistent level in the RGS data. 
The start times of bursts in the RGS data are consistent with the start times in the EPIC/pn data, when bursts are visible in both datasets.  The burst durations are naturally longer in the RGS data, which covers a softer energy band than the EPIC/pn data.  Burst durations in the RGS data range from 22 to 290 seconds.  
The peak intensity ranges from 1.6 to 20.5 counts s$^{-1}$ with an average peak intensity of 9.6 counts s$^{-1}$.  The local persistent count rate varies from 4.8 counts s$^{-1}$, down to 0.2 counts s$^{-1}$ during dips. 
We recorded a cumulative exposure time of 8555 seconds for the 68 bursts defined using the RGS data alone.  It should be noted that while this burst exposure time is a little more than twice the burst exposure time of the 2000 observations, since CCD 7 of the RGS1 was lost after the 2000 data were acquired the effective exposure in the critical $13\, {\rm \AA}$ band is comparable between the two data sets.    
\clearpage
\clearpage
\section{Burst Spectra}

We started, as we did in CPM, by extracting the first-order RGS spectra for each of the bursts identified in the RGS lightcurve using start and stop times defined by the RGS data.  
For our initial analysis we combined all bursts, treating the two bursts in double burst series and the three bursts in triple burst series as individual bursts.  We combined the data within each of the seven observations 
and generated a single, representative response matrix for each of the seven epochs.  Background spectra were generated for each observation using spatially offset regions according to the standard SAS routines.  
The spectral files and response matrices for each of the seven observations were then combined using the SAS tool {\it rgscombine}, and fit using XSPEC version 11 \citep{XSPEC}.  We verified that the results of fitting the seven individual spectra simultaneously and fitting the single combined spectrum were the same.  
Using discrete features from known transitions in the persistent spectrum we confirm that the wavelength scale is accurate to $\sim 10\, {\rm m\AA}$.  According to \citet{RGS}, the effective area of the RGS is calibrated to an accuracy of $\sim 5\%$ for wavelengths above $\sim 8\, {\rm \AA}$.  
Deviations at the long wavelength end discovered more recently are largely corrected in SAS version 7.0.0.  The remaining uncertainty in the effective area is $\leq 10\%$ at the longest wavelengths (RGS instrument team, private communication).  
As we are concerned with the discrete spectral structure and not the exact shape of the continuum spectrum, this does not significantly impact our analysis.  In the data we present here, the background rate becomes a significant fraction of the source rate longward of $\sim 34\, {\rm \AA}$.  We therefore conservatively consider data only in the wavelength range $\lambda = 8-32\, {\rm \AA}$.   

\begin{figure*}[htb]  
\includegraphics[angle=90,height=4.5in]{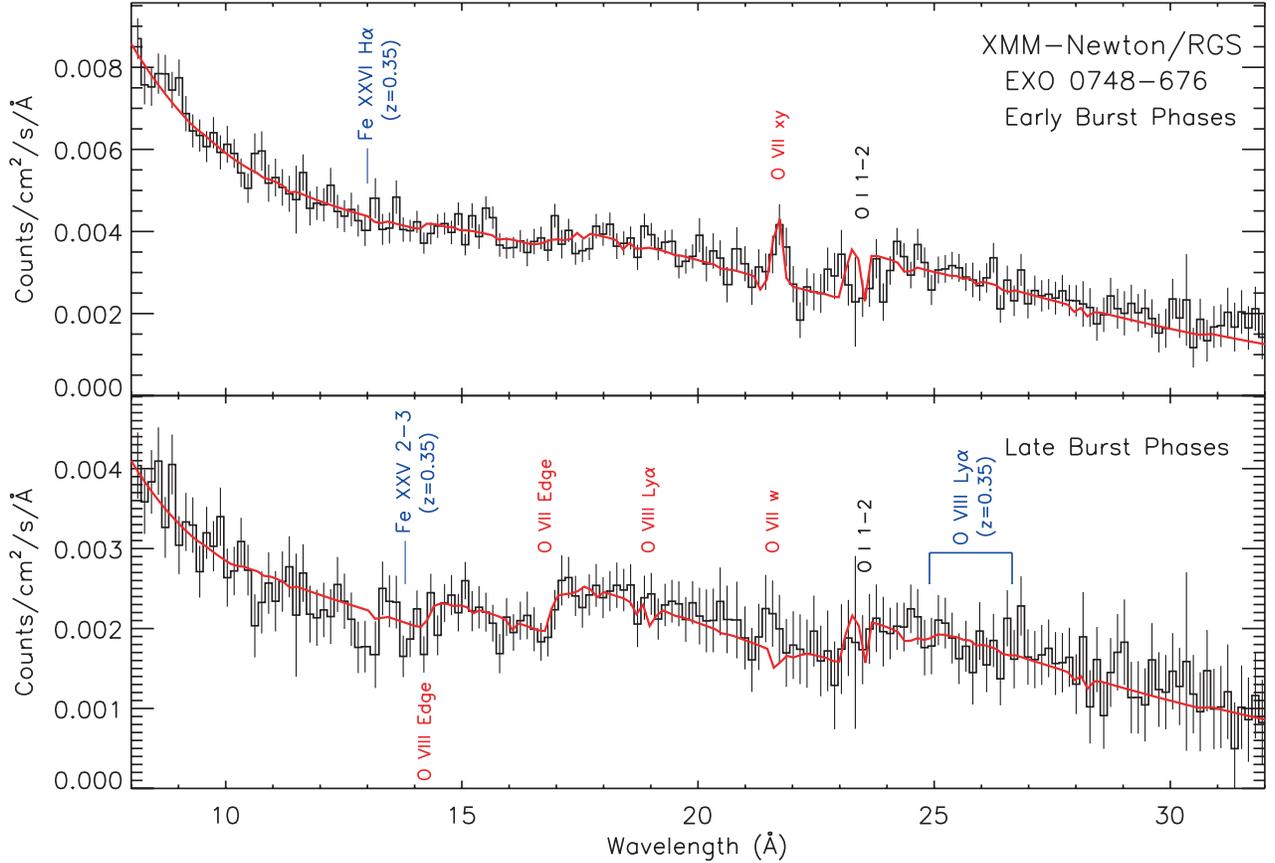} 
\caption{Background-subtracted fluxed spectra from early and late phases of the average EXO 0748-676 burst in the 2003 data.  Data from 67 type-I X-ray bursts are combined.  Early and late phases are defined using start and stop times derived from the RGS lightcurve.  Data are binned to $\Delta \lambda = 0.148\, {\rm \AA}$.  An empirical continuum model, modified by circumstellar absorption, is superimposed in red.  The model is generated by fitting the observed \ion{O}{7} features and synthesizing absorption structure for other abundant elements assuming an ionization parameter of $\xi = 10$.  The most prominent circumstellar features are labelled in red.  Interstellar O I absorption is labelled in black.  The positions of the neutron star photospheric absorption features identified in \citet{CPM} are indicated in blue.}   
\vspace{0.00in}
\label{fig2} 
\end{figure*}

As in CPM, we separated the data from each burst into early and late phases in order to account for spectral evolution over the course of the burst.  As in our earlier work, we experimented with different ways of defining the 'early' and 'late' phases (for instance, attempting to have approximately equal numbers of counts in both).  In view of the fact that the signal to noise ratio in the spectrum is necessarily limited, the differences between the pairs of spectra extracted using different definitions are not significant.  In this analysis we therefore simply divided each burst in half by duration as we did in CPM.  For display purposes we generated spectra for the early and late phases of the average burst by combining all the data from both RGS instruments using the SAS tool {\it rgsfluxer}.  The background-subtracted spectra are shown in Figure \ref{fig2}.  
\clearpage
\clearpage
The dominant spectral features are generated in highly ionized circumstellar material.  We fit the early and late phase spectra with the circumstellar absorption model described in CPM.  For every ion, the model self-consistently accounts for absorption in all transitions arising from a given level, out to the photoelectric continuum limit \citep{Sako}.  
We use an empirical continuum model that consists of black body and power-law emission with neutral interstellar absorption.  We first fit the \ion{O}{7} emission and absorption structure in both spectra.  As in the 2000 data analyzed in CPM, a broad \ion{O}{7} emission feature is evident in the early phases of the bursts. This is dominated by the \ion{O}{7} $n=1-2$ intercombination line (\ion{O}{7} $x,y$) at a rest wavelength of $21.80$ \AA.  The emission line disappears in the late phases of the bursts as the depth of the \ion{O}{7} absorption edge increases.    
The intensity of the \ion{O}{7} features in these data is much weaker than in the 2000 data; the equivalent width in the \ion{O}{7} line is $EW \sim 3\, {\rm eV}$ in the early phases of the bursts compared to $EW \sim 6\, {\rm eV}$ in the early phases of the 2000 spectra, and the ion column density is $N_{\rm O VII} \sim 4\times 10^{17}\, {\rm cm^{-2}}$ in the early phase and $N_{\rm O VII} \sim 1\times 10^{18}\, {\rm cm^{-2}}$ in the late phases of the burst, compared to $N_{\rm O VII} \sim 8\times 10^{17}\, {\rm cm^{-2}}$ and $N_{\rm O VII} \sim 2.5\times 10^{18}\, {\rm cm^{-2}}$ in the early and late phases of the 2000 bursts.  
The turbulent velocity in the absorbing material, which is constrained by the equivalent width of the \ion{O}{7} resonance absorption line ($w$) at 21.60 \AA, is much lower in these data than in the 2000 data;  we measure 
effectively upper limits to turbulent velocities of $v_{\rm t} \la 20$ km s$^{-1}$ in the early phase and $v_{\rm t} \la 10$ km s$^{-1}$ in the late phase of the bursts, compared to positive detections of $v_{\rm t} \sim 100$ km s$^{-1}$ and $v_{\rm t} \sim 200$ km s$^{-1}$ in the 2000 data.    

As in CPM, to fully account for any additional absorption structure due to circumstellar material, we synthesized the absorption spectra for the hydrogen- and helium-like ions of carbon, nitrogen, neon, magnesium, silicon, and the L-shell ions of iron, based on the parameters measured at oxygen.  
To estimate the fractional abundances for each ion we use an ionization parameter of $\xi = L/n_{\rm e}r^2 \sim 10$, which we estimate is an upper limit based on the lack of observed \ion{O}{8} emission.  Using a lower ionization parameter will only decrease the fractional abundance of all these ions except perhaps He-like carbon.  We assume solar abundance ratios.  We then scaled the turbulent velocity for each ion from the value measured at \ion{O}{7} assuming a constant temperature for all ions.  The full circumstellar model is superimposed on the spectra shown in Figure \ref{fig2}.    

After accounting for the circumstellar contribution to the observed spectra we search for the absorption features from the neutron star photosphere that were identified by CPM in the 2000 data, or any other significant absorption.   
The locations of the features we previously identified, $13.0\, {\rm \AA}$ in the early phases of the burst, and $13.8$, $25.2$ and $26.4\, {\rm \AA}$ in the late burst spectra are indicated in Figure \ref{fig2}.  
This spectral structure is not repeated here.  
The simplest nontrivial explanation for this non-detection may be our particular data selection.  Since the EPIC data were largely not available during the 2000 observations we do not know what selection biases may have been introducted in that analysis.
We therefore explored alternative selection criteria for the 2003 data, particularly our treatment of double and triple bursts. For example, we extracted spectra selecting only the single bursts and the first of double and triple bursts, which the analysis of \citet{Boirin} suggests involve different burning than the second and third bursts in such series.  
We also explored the effect of using EPIC/pn defined burst start and stop times instead of the RGS defined times.  None of these changes in selection criteria had a significant effect on the resulting spectra.   

The only evidence of absorption structure that can not be attributed to the circumstellar material is a marginal absorption feature at $\lambda = 13.0\, {\rm \AA}$ in the late phases of the bursts.  
We find that this feature has the highest equivalent width when we define the start and stop times using the EPIC/pn lightcurve, which are shorter than the RGS-defined times and therefore minimize the contribution from the persistent spectrum.  
Using the EPIC/pn defined times we estimate a significance by fitting a simple Gaussian to the line.  We measure a significance of $3 \sigma$ relative to a zero amplitude fit, and an equivalent width of $EW = 0.09 {\pm 0.025}\, {\rm \AA}$.  There is no evidence for an absorption feature at $\lambda = 13.0\, {\rm \AA}$ in the persistent spectrum.      

\section{Discussion}

The 2003 burst spectra show different neutron star photospheric absorption structure than the spectra reported in CPM.  In these data we do not see the absorption feature at $13.0\, {\rm \AA}$ that we identified as the gravitationally redshifted $n=2-3$ transition of \ion{Fe}{26} in the early phases of the bursts, and we do not see the feature at $13.75\, {\rm \AA}$ that we identified as the gravitationally redshifted $n=2-3$ transition in \ion{Fe}{25} in the late phases of the bursts. Either our initial detections were due to highly improbable statistical fluctuations, or the conditions in the neutron star photosphere have changed.  

\begin{figure*}[htb] 
\includegraphics[angle=90, height=2.3in]{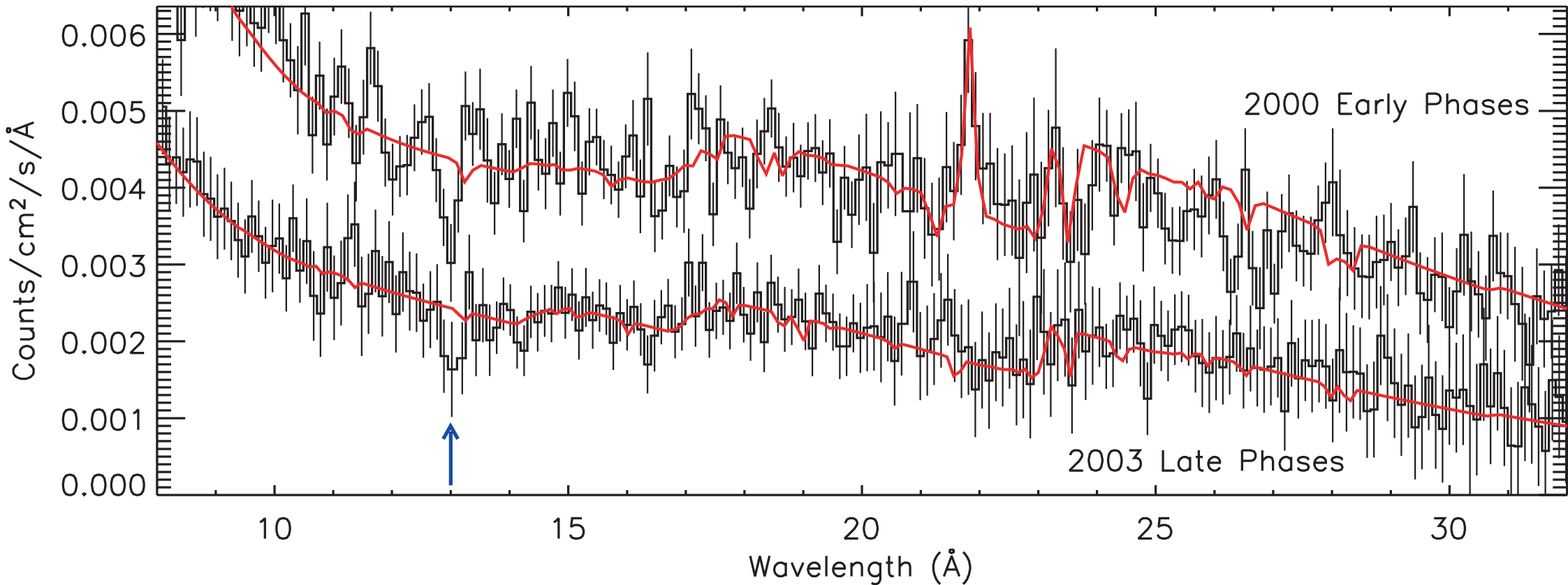} 
\vspace{-0.0cm}
\caption{Comparison of the RGS spectra from the early phases of the 2000 bursts and the late phases of the 2003 bursts.  The top histogram represents the 2000 data, and the bottom histogram represents the 2003 data.  The circumstellar model for each data set is superimposed in red.  The 2000 data are reproduced from \citet{CPM}, offset by a constant $0.001$ counts cm$^{-2}$ s$^{-1}$ \AA$^{-1}$ for visual clarity.  The 2003 spectrum was generated using start and stop times derived from the EPIC/pn lightcurve.  The location of the redshifted \ion{Fe}{26} H$\alpha$ absorption feature identified in the early phases of the 2000 burst data is indicated with a blue arrow.} 
\label{fig3} 
\end{figure*}

We see no absorption features in the early phases of the bursts, and a marginal absorption feature at $13.0\, {\rm \AA}$ in the late phases of the bursts.  This feature is not particularly statistically significant, and as a single spectroscopic feature it would be difficult to conclusively identify.  
However, it is at exactly the same wavelength as the feature in the early phases of the 2000 data set that we identified as gravitationally redshifted $n=2-3$ transition of \ion{Fe}{26}.  This coincidence is illustrated in Figure \ref{fig3}.  
The signal to noise ratio of this feature, $3 \sigma$ relative to a zero amplitude fit, is half of the value, $7 \sigma$, for the corresponding feature in the early phases of the 2000 burst data.  We measure an equivalent width of $0.095\, {\rm \AA}$, compared to a value of $0.13\, {\rm \AA}$ in the 2000 data.  
The difference is due at least in part to a significant contribution of the persistent emission to the total flux during the 2003 observations.  This was negligible in the 2000 data.  
We have not attempted to subtract the average persistent RGS spectrum from the average burst spectra, because the shape of the spectrum in the RGS band is very sensitive to the precise properties of obscuring and absorbing material, which are clearly strongly variable. If the properties of this material during the ensemble of burst intervals is in some way not statistically identical to those averaged over the entire exposure, by subtracting the persistent spectrum we could be introducing artefacts instead of removing an unrelated spectral component (see \citet{Boirin,AD2006}).   
\clearpage
\clearpage

We briefly discuss some possible physical effects that could have changed the average photospheric spectrum between 2000 and 2003. 
In CPM we interpreted the changing spectral structure in terms of a changing ionization balance in the photosphere.  In the early hot phases of the burst only the \ion{Fe}{26} ions were sufficiently abundant in the photosphere to generate the observed absorption feature at $13.0\, {\rm \AA}$.  
As the photosphere cooled, the ionization balance shifted such that \ion{Fe}{25} became dominant.  If we can identify the $13.0\, {\rm \AA}$ feature in the late phases of these new data as the redshifted $n=2-3$ transition in \ion{Fe}{26} by its coincidence with the feature in the early phases of the 2000 data, then these spectra could be interpreted similarly.  
In these new data the lack of absorption features in the early phases of the bursts, and the possible presence of the $13.0\, {\rm \AA}$ feature in the late phases suggests that the ions are fully stripped in the hot early phases of the burst and that the population of \ion{Fe}{26} ions only becomes sufficiently abundant in the late phases as the bursts cool.  
The changes in the observed spectra from 2000 to 2003 could then be understood as an overall shift in the ionization conditions in the neutron star photosphere.
We would then expect to observe a gravitationally redshifted line from \ion{Fe}{25} at still later burst times.  Unfortunately, the persisent source emission rapidly dominates the burst emission, and we are unable to distinguish the burst spectrum at later times. 

We do indeed expect the absorption structure in the burst spectra to be extremely sensitive to changes in the photospheric conditions. Inspection of just the Saha ionization balance for Fe in the relevant regime shows that the fractional abundance of H- and He-like Fe is extremely sensitive to the local electron temperature and density. In addition, the relative population of $n=2$ is also very sensitive to the details of radiative transfer in the transitions between the lowest several levels (the radiative transition probability $n = 2 \rightarrow 1$ in highly ionized Fe is so high that even at the densities expected in neutron star atmospheres at gas temperatures of order 1 keV, collisions cannot enforce local thermodynamic equilibrium between the lowest levels and the ground state; \citet{Bildsten,Chang05}; Paerels et al.\ (2007, in preparation); Lanz et al.\ (2007, in preparation). 

We therefore investigate what independent evidence exists that indicates that average photospheric conditions did, or did not change between the 2000 and 2003 datasets. Obviously, a measurable change in the average burst continuum shape would be a strong indicator of a change in the effective temperature, and hence very likely in the characteristic electron temperature in the photosphere.
For this, we need the EPIC spectra, since RGS sensitivity cuts off at photon energies above 2 keV. Unfortunately, the 2000 data were acquired during the commissioning and calibration phases of the mission, so the instrument configurations were not optimized for burst spectroscopy.  
Only two of the EXO~0748$-$676 observations in 2000 have simultaneous EPIC coverage, and only part of one observation was acquired with the EPIC/pn in small window mode, such that the spectra are not piled-up during the bursts.  We identified three bursts in these data, which we can compare to the 2003 bursts as described in \citet{Boirin}.  
The durations of these bursts, 135, 178, and 52 seconds, are consistent with the range of durations of the bursts in the 2003 data.  The peak intensity in the $5$ to $10\, {\rm keV}$ band, $86, 87$ and $51$ counts s$^{-1}$ are also consistent with the values for the 2003 bursts.  
We analyzed the temperature evolution of the two bursts that did not occur during a dip.  Following the methods of \citet{Boirin} we measure a black body temperature of $kT = 1.8\, {\rm keV}$ at the peak of each burst.  Both bursts then decay to a temperature of $kT \sim 1.3\, {\rm keV}$.  
This is consistent with the temperature evolution of the 2003 bursts.  However, all three of these bursts occurred during the last $9\, {\rm ks}$ of the 2000 data.  Since the full 2000 dataset was acquired over the course of two months, these three bursts are not necessarily representative of the conditions throughout the 2000 data set.  
There are indications from RXTE data (M. Krauss, private communication) that the peak and average burst temperature in EXO~0748$-$676 have varied considerably over the lifetime of the RXTE mission, with variations as large as $\pm 0.5\, {\rm keV}$ over the course of a few months.  
Unfortunately we do not have simultaneous RXTE measurements during the 2000 XMM-Newton observations.  On the other hand, comparison of the RGS burst parameters during the 2000 and 2003 observations is straightforward, and the distributions of peak intensity and duration as measured by the RGS, are consistent between the two data sets.  
But this comparison is less instructive, since the RGS data are affected by (highly variable) obscuration and absorption by the circumstellar material.  

Evidence for a change in the effective temperature during bursts is therefore inconclusive. But factors other than the photospheric temperature could affect the appearance of the photospheric absorption spectrum. A change in the average density at which the absorption spectrum is formed would have an equally strong effect on the overall ionization balance, and through the associated shift between the collisional and radiative rates, which affects both the ionization and the excitation balances. And in fact, simply changing the distribution of Fe through the atmosphere could change the absorption spectrum. Unfortunately, we have no independent measure of the average photospheric density (the discrete absorption spectrum is in fact the best indicator). The broad band shape of the continuum is of course weakly sensitive to the average photospheric density, but, again, we do not have the necessary EPIC spectroscopy to systematically compare continuum shapes in 2000 and 2003.   

\citet{Bildsten} pointed out that if the measured gravitational redshift of $z = 0.35$ is in fact correct, then the neutron star is smaller than the radius of the innermost stable Keplerian orbit, and the accretion disk effectively terminates in vacuo, with the accreting material falling nearly radially into the atmosphere. The result is that heavy nuclei are spallated in the atmosphere, and the distribution of any given element through the atmosphere is inhomogeneous. Should there be any change in the energy with which the nuclei fall into the atmosphere, then this stratification would also change.  

The accretion flow does indeed appear to have changed in some respects: the circumstellar conditions in the persistent state of the source have changed considerably from 2000 to 2003 (a detailed comparison of the persistent spectra will be presented in van Peet et al.\ (2007, in preparation).  
It is immediately evident from the RGS lightcurve in Figure \ref{fig1} that the scale height of the circumstellar material, which mostly obscured and reprocessed the soft central emission in the 2000 data, is significantly diminished in 2003. 
The central emission now dominates the soft X-ray lightcurve. At the same time, however, the luminosity of the accretion disk appears not to have changed appreciably.   
Using the hard band EPIC/pn data, which is less sensitive to variability in the obscuring circumstellar material, we find an unabsorbed flux of $\sim 2.8 \times 10^{-10}$ erg cm$^{-2}$ s$^{-1}$ from 0.6 to 10 keV \citep{Boirin} for the 2003 data with maximum variations of $30\%$ over the course of all observations.  
\citet{Homan} report an unabsorbed flux of $\sim 9 \times 10^{-11}$ erg cm$^{-2}$ s$^{-1}$ from 5 to 10 keV for the 2000 data, which is equivalent to $\sim 2.8 \times 10^{-10}$ erg cm$^{-2}$ s$^{-1}$ in the 0.6 to 10 keV band for their spectral model.  Although we note that the source luminosity is not necessarily a direct measure of the local accretion conditions, we conclude that the basic properties of the inner accretion disk have probably not changed much: the mass transfer rate is likely comparable between the 2000 and 2003 epochs, and the inner edge of the disk has likely not moved in radius (which of course it would indeed not be expected to do if it terminates at the innermost stable circular orbit). 
 
\section{Conclusion}

We have analyzed soft X-ray spectroscopic observations with XMM-Newton/RGS of the X-ray bursts from the LMXB EXO0748$-$646, obtained during the fall of 2003. The observations were performed to corroborate the detection of narrow soft X-ray atomic absorption lines in the burst spectrum as observed in 2000, in a spectrum of comparable depth. We accumulated a spectrum over 68 bursts, occurring in singles, doubles, and triples, for a total of 8555 seconds net RGS burst exposure time.
The $8-35$ \AA\ burst spectrum in 2003 appears to be featureless, even when split into 'early' and 'late' burst phase spectra, with the possible exception of a marginal feature at $\lambda = 13.0$ \AA\ during the late burst phases; this latter feature attracts attention only because it appears at the same wavelength at which the putative \ion{Fe}{26} $n=2-3$ transition was seen in 2000, during the early burst phases. Whatever measures of the photospheric continuum shape we have available for comparison between the two epochs appear to indicate that the average burst properties have not changed dramatically, but this conclusion is based on only three 2000 bursts observed with EPIC. 
Given the fact that the photospheric discrete absorption spectrum is very sensitive to the precise conditions in the atmosphere, it remains possible that a change in average burst parameters has reduced the contrast in the absorption lines.



\acknowledgments

Our research is based on data obtained with XMM-Newton, an ESA science mission with instruments and contributions funded directly by ESA member states and the USA (NASA). 
We thank Fred Jansen for providing the XMM-Newton Director's Discretionary time for these observations. 
FP gratefully acknowledges support from NASA under grants NAG5-7737 and NNG05GG09G (NASA ATP Program). \\ 



Facilities: \facility{XMM} 





\clearpage





\clearpage


\end{document}